# Soft phonons reveal the nematic correlation length in Ba(Fe$_{0.94}$Co$_{0.06}$)$_2$As$_2$

F. Weber[1], D. Parshall[2,3], L. Pintschovius[1], J.-P. Castellan[1,4], M. Kauth[1], M. Merz[1], Th. Wolf[1], M. Schütt[5], J. Schmalian[1,6], R. M. Fernandes[5] and D. Reznik[3,7]

[1] Institute for Solid State Physics, Karlsruhe Institute of Technology, 76021 Karlsruhe, Germany
[2] NIST Center for Neutron Research, National Institute of Standards and Technology, Gaithersburg, Maryland 20899, USA
[3] Department of Physics, University of Colorado at Boulder, Boulder, Colorado, 80309, USA
[4] Laboratoire Léon Brillouin (CEA-CNRS), CEA-Saclay, F-91911 Gif-sur-Yvette, France
[5] School of Physics and Astronomy, University of Minnesota, Minneapolis, Minnesota, 55455, USA
[6] Institute for Theoretical Condensed Matter Physics, Karlsruhe Institute of Technology, 76128 Karlsruhe, Germany
[7] Center for Experiments on Quantum Materials, University of Colorado at Boulder, Boulder, Colorado, 80309, USA

**Abstract**
Nematicity is ubiquitous in electronic phases of high-T$_c$ superconductors, particularly in the Fe-based systems. While several experiments have probed nematic fluctuations, they have been restricted to uniform or momentum averaged fluctuations. Here, we investigate the behavior of finite-momentum nematic fluctuations by utilizing the anomalous softening of acoustic phonon modes in optimally doped Ba(Fe$_{0.94}$Co$_{0.06}$)$_2$As$_2$. We determine the nematic correlation length and find that it sharply changes its T-dependence at T$_c$, revealing a strong connection between nematicity and superconductivity.

In several correlated quantum materials the high-temperature crystal structure changes due to the self-organization of electrons into a state with lower rotational symmetry [1]. Given the analogous symmetry breaking in liquid crystals, such electronic phases are called nematic [2]. In copper oxide superconductors, they may be associated with the formation of charge stripes whose role is not yet clearly established [3-6]. In Fe-based superconductors, nematic order is well supported in many families of the iron pnictides [7,8] and chalcogenides [9]. It can be viewed as due to partially melted striped spin density-waves, that are often observed to order slightly below the nematic transition [10-12]. Alternatively, this symmetry lowering has been proposed to be due to ferro-orbital ordering [13-15].

In the widely investigated material Ba(Fe$_{1-x}$Co$_x$)$_2$As$_2$ magnetic fluctuations, associated with the spin density-wave ground state, break the tetragonal symmetry in the *a-b* plane above the magnetic ordering temperature and introduce an orthorhombic distortion via a magnetoelastic coupling [16]. The coupling to anisotropic magnetic fluctuations shortens atomic bond lengths along one direction and lengthens them along the other [17-21]. It has been shown that this orthorhombicity of the atomic lattice is driven by electronic nematic order [22] with strong evidence for magnetic fluctuations causing this new state of matter [23].

Nematic order triggers orthorhombic order. Hence, the nematic order parameter $\varphi$ has zero wave vector ($q = 0$) and is proportional to the shear distortion $\epsilon_{66}$ [24,25] due to the nemato-elastic coupling $\lambda\varphi\epsilon_{66}$ in the free energy. In the para-nematic phase, nematic fluctuations soften the shear modulus $C_{66}$. The same coupling then yields the general relation

$$C_{66}(q) = \frac{C_{66}^0}{1+\frac{\lambda^2}{C_{66}^0}\chi_{nem}(q)} \qquad (1)$$

between the elastic constant, the bare elastic modulus $C_{66}^0$ without nematicity and the nematic susceptibility $\chi_{nem}$. Eq.(1) was derived in Ref. [24] for $q = 0$. The generalization to finite momenta is straightforward. $C_{66} \equiv C_{66}(q = 0)$ has been measured either directly by resonant ultrasound or indirectly via the Young's modulus $Y_{110}$ in three-point bending setups [24,26,27]. These measurements have shown that $q = 0$ nematic fluctuations in Ba(Fe$_{1-x}$Co$_x$)$_2$As$_2$ are present for a wide temperature and doping range, including nematic quantum criticality near optimal doping. At chemical substitution levels for which the tetragonal-to-orthorhombic transition is suppressed, these fluctuations generally increase on cooling but decrease rapidly on entering the superconducting phase, signaling a competition between superconductivity and nematic order [24]. The uniform nematic susceptibility has also been probed via electronic Raman scattering [28,29] and elasto-resistance [30], while the momentum averaged nematic response affects the NMR and NQR line broadening [31]. An essential open problem is the behavior of nematic fluctuations at small but finite wave vectors $q > 0$. What is the size of fluctuating nematic regions and how do these dynamic nematic droplets change character upon entering the superconducting state?

In this paper, we show that the soft transverse acoustic (TA) phonons, with wavelength of up to 25 unit cells, dispersing along the [010] direction and polarized along [100], serve as efficient probes of spatial nematic correlations. These TA phonons correspond to the vibrational shear modes $\epsilon_{66}$ of the whole crystal in the long wave length limit and, hence, correspond to the soft mode of the structural/nematic phase transition in Ba(Fe$_{1-x}$Co$_x$)$_2$As$_2$. Our conclusion utilizes the generalized relation

$$E(q) = \sqrt{\frac{C_{66}(q)}{\rho}}|q| \qquad (2)$$

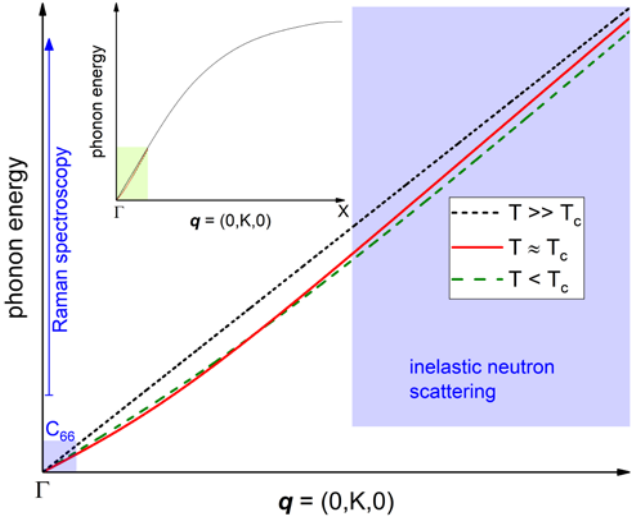

**FIG. 1.** Schematic of the temperature dependence of the soft phonon mode dispersion observed in Ba(Fe$_{0.94}$Co$_{0.06}$)$_2$As$_2$. The inset displays the full dispersion from the zone center, Γ, to the zone boundary, $X$; the main panel zooms the small-momentum green-shaded area. Shown are the characteristic evolution of the TA dispersion described by Eq. (4) from high temperatures, $T \gg T_c$, down to the superconducting transition temperature $T_c$ and below, $T < T_c$. The values of $\xi$ used to generate the curves were taken from the fits of Fig. 4e at the corresponding temperatures. The blue vertical arrow and shaded areas indicate the energy-momentum range probed by Raman spectroscopy ($\boldsymbol{q} \approx 0, \omega > 0$), measurements of the elastic constant $C_{66}(0)$ ($\boldsymbol{q} \approx 0, \omega = 0$) and our inelastic neutron scattering investigation ($\boldsymbol{q} \neq 0$).

for the phonon dispersion near the zone center $\Gamma$ with mass density $\rho$ and momentum-dependent elastic modulus $C_{66}(\boldsymbol{q})$. The latter is related via Eq.(1) to the momentum-dependent nematic susceptibility

$$\chi_{nem}(\boldsymbol{q}) = \frac{\chi_{nem}(\boldsymbol{q}=0)}{1+\xi^2 q^2} \qquad (3)$$

with nematic correlation length $\xi$. Thus, for systems near a nematic instability, where $C_{66}(\boldsymbol{0})$ is small, one expects a linear phonon dispersion only on a very narrow momentum range near $q = 0$, with non-linear behavior setting on around $|\boldsymbol{q}| \sim \xi^{-1}$.

Combining equations (1)-(3) yields the dispersion relation of the TA phonon, including the effects of nematic fluctuations

$$E(q) = \sqrt{\frac{C_{66}^0}{\rho\left(1+\frac{\lambda^2 \chi_{nem}(q=0)}{C_{66}^0(1+\xi^2 q^2)}\right)}} q^2. \qquad (4)$$

We measured the TA phonon dispersion of optimally doped Ba(Fe$_{0.94}$Co$_{0.06}$)$_2$As$_2$ via inelastic neutron scattering (INS). The experiments were performed on the 4F2 cold triple-axis and on the 1T thermal triple-axis spectrometers at the ORPHEE reactor (Laboratoire Leon Brillouin (LLB), Saclay, France). The compositions of self-flux grown single crystals of Ba(Fe$_{1-x}$Co$_x$)$_2$As$_2$ grown at the Institute for Solid State Physics (KIT) were determined by single-crystal x-ray diffraction and energy-dispersive x-ray spectroscopy as done in previous work in our institute [32,33]. Here, wave vectors are given in reciprocal lattice units (r.l.u.) of ($2\pi/a, 2\pi/a, 2\pi/c$) where $a$ and $c$ are the lattice constants of the tetragonal unit cell. For more experimental details, please see the Appendix.

The main conclusions of our paper are summarized in Figure 1, where we indicate the relevant energy and frequency regimes of different experimental methods. At high temperatures, where the nematic correlation length is presumably small, the TA phonon branch shows the expected linear dispersion relation $E(\boldsymbol{q}) = \sqrt{\frac{C_{66}(\boldsymbol{0})}{\rho}} |\boldsymbol{q}|$ indicated by the straight dotted line. The slope of the dispersion is proportional to $\sqrt{C_{66}(\boldsymbol{0})}$ and, thus, determined by the uniform nematic susceptibility through Eq.(1). For low temperatures, where the nematic correlation length is presumably large, this linear behavior is expected to be confined to very small momenta. These momenta are well below the momentum range accessible in INS experiments, which probe the energy-momentum range indicated by the large blue-shaded box in Figure 1. Variation of the nematic correlation length with temperature results in different temperature dependences of the elastic modulus (at $q = 0$) and acoustic phonons at finite wave vectors accessible to the INS experiments. We will show that by combining the phonon and elastic modulus data we can determine the temperature evolution of $\xi(T)$.

Figure 2 illustrates how the softening of long-wave-length acoustic phonons appears in the INS data. We performed constant energy scans along (2,$K$,0), $-0.1 \leq K \leq 0.1$ (the scan geometry is displayed in the inset of Fig. 2). Two peaks on opposite sides of $K = 0$ correspond to acoustic phonons propagating along $K$ and along $-K$. The asymmetric line shape is due to the tilted resolution ellipsoid and corresponding focusing effects intrinsic to the triple-axis instrument [34]. Because in our measurements the resolution function also detected a small contribution from longitudinal acoustic phonons centered at $K = 0$ and possibly a tail from the nearby Bragg reflection, our constant energy scans were fit with three peaks. The observed phonon softening is manifested as an increased separation between the peaks, since the scan at a fixed energy crosses the dispersion curves at larger $|K|$ (see inset of Fig. 2).

It is established from thermodynamic experiments that the strongest softening of $\sqrt{C_{66}(\boldsymbol{0})}$ occurs at $T = T_c$, followed by a hardening on further cooling below $T_c$ [26,27] (dashed line in Fig. 3). In contrast to this behavior of $\sqrt{C_{66}(\boldsymbol{0})}$, INS data reveal further phonon softening on cooling below $T_c$. Figure 3 shows that the phonon at $E = 1$ meV first softens from 300 K to about 70 K. Below 70 K the temperature dependence is nearly flat down to $T_c$ where the softening resumes in the superconducting state. Results obtained for larger energy transfers of $E = 1.5$ meV and 2.5 meV show qualitatively the same behavior (see Appendix).

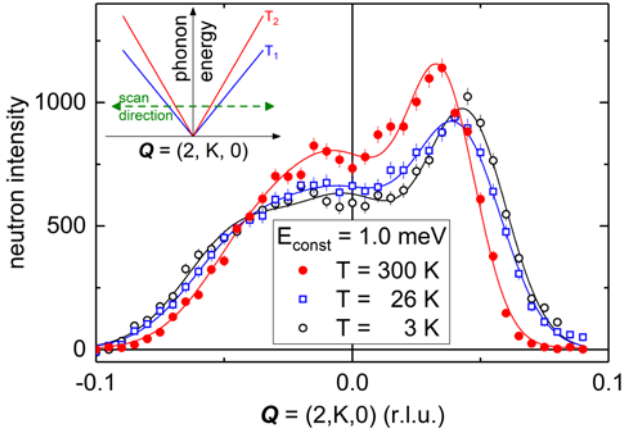

**FIG. 2:** Inelastic neutron scattering data of the transverse acoustic mode in Ba(Fe$_{0.94}$Co$_{0.06}$)$_2$As$_2$ dispersing along the [010] direction and polarized parallel to the [100] direction at three temperatures well above, at and below $T_c = 25$ K. Measurements were done at fixed neutron energy transfer of $E = 1.0$ meV. Estimated flat experimental backgrounds were subtracted and intensities corrected for the Bose factor. Shift of the peak position to higher values of $K$ corresponds to softening. Solid lines are fits to the data as described in the text. The inset shows a schematic view of the inelastic scans for the case of an acoustic phonon branch with a temperature dependent slope at two temperatures $T_1 < T_2$.

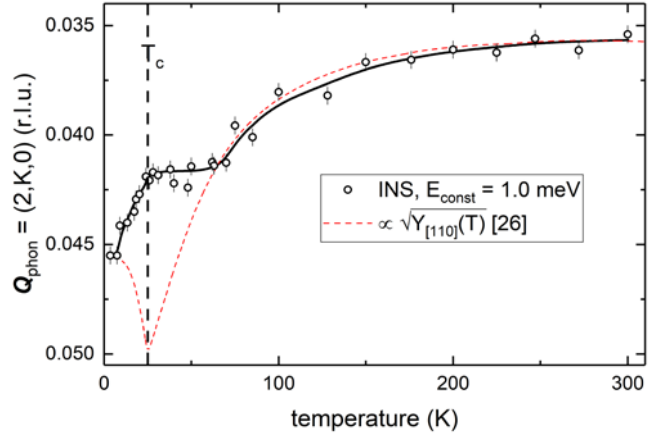

**FIG. 3.** Temperature dependence of the phonon peak position at $Q_{phon} = (2, K, 0)$ (circles) observed at $E = 1$ meV. Solid line is a guide to the eye. For a more intuitive reading, we plot the $K$ scale upside-down because the shift of the peak position to higher values of $K$ corresponds to a softening. The dashed (red) line represents the naïve behavior expected based on measurements of $Y_{[110]}$ linearly scaled to the observed phonon softening at $T \geq 90$ K [26].

At first glance, the softening observed here seems to contradict the shear modulus hardening previously observed. However, as explained above, the non-linear dispersion of the phonon becomes more pronounced as temperature is lowered and the nematic susceptibility is enhanced. As we will show quantitatively below, the INS and shear modulus measurements are perfectly consistent. Moreover, our INS data allows us to extract the characteristic nematic correlation length $\xi$ based on fits to the phonon dispersion $E(q)$ using Eq. (4). Note that our previous investigation in *undoped* BaFe$_2$As$_2$ and SrFe$_2$As$_2$ found a softening of the TA phonons on approaching the nematic phase transition [35], which tracks the nematic and magnetic fluctuations in the tetragonal phase [36].

Figures 4(a)-(d) show the phonon dispersion points $E(q)$ obtained from our experiments along with fits of Eq.(4) to the data. Because of the relation $\lambda^2 \chi_{nem}(q = 0) = C_{66}^0 (C_{66}^0 / C_{66} - 1)$ [24], the nematic correlation length $\xi$ is the only remaining fitting parameter since $C_{66}^0$ and $C_{66}$ for the given temperature are known from measurements of the elastic moduli [26] (see Appendix for more details).

The experimental dispersion stays quasi linear within the experimental uncertainty down to about 100 K [Figs. 4(a)(b)], implying a short correlation length of $\xi \leq 10$ Å. Results for the fitted linear coefficient, $C_{66}$, extracted from a linear fit to phonon energies (assuming $\xi = 0$), are shown for temperatures less than 100 K as circles in the inset of Figure 4(e). At lower temperatures the phonon data are not in the linear dispersion regime, as shown in Figs. 4(c)-(d) by contrasting the solid lines with the dashed and dotted lines. In these cases, we used $C_{66}(q = 0, T)$ from thermodynamic measurement of the Young's modulus $Y_{110}$ [26] performed on samples grown in the same lab to constrain our fits. Because $Y_{[110]}$ does not give the absolute values for $C_{66}(q = 0, T)$, they were scaled to fit the deduced values from INS for temperatures $T \geq 90$ K [red line and circles, respectively, in the inset of Fig. 4(e); see also Appendix].

Fitting results treating $\xi$ as a free parameter are shown in Figure 4(a)-(d) for selected temperatures. Figure 4(e) shows that $\xi(T)$ obtained via such fits displays a sharp peak below 100 K with a maximum at $T_c$. It is the reduction of $\xi$ below $T_c$ that promotes the observed phonon softening at intermediate momenta measured by INS, stemming from the non-linearity of the phonon dispersion. The weak temperature dependence of the phonon for $T_c \leq T \leq 70$ K [Fig. 3] is, in fact, due to the strong increase of the non-linear term in the dispersion relation [Fig. 4(e)].

It is important to rule out other more trivial sources for the non-linear corrections to the phonon dispersion observed here. While the coupling between TA phonons and electrons vanishes in the simplest form of the electron-phonon Hamiltonian, Umklapp processes and disorder effects promote a non-zero coupling. As a result, in a non-interacting electron gas, this self-energy correction to the phonon dispersion is proportional to the standard Lindhard function $\Pi$. At temperatures well below the Fermi temperature, as it is presumably the case near $T_c$, the Lindhard function depends on the ratio $E(\mathbf{q})/\mathbf{q}$, implying that the phonon dispersion remains linear. Thus, the observed non-linear effect does not arise from the coupling to free electrons. Instead, the above general expression Eq.(3) of the finite momentum nematic susceptibility offers the most natural explanation for the observed non-linearity.

The observed maximum of $\xi$ at $T_c$ also constrains the possible microscopic mechanisms behind the formation of the nematic state. In the spin-driven nematic scenario, nematic fluctuations are composite magnetic fluctuations. As a result, not only the nematic and magnetic transition temperatures are related, but also the nematic and antiferromagnetic correlation lengths, $\xi$ and $\xi_{AFM}$. Following Ref. [37], we find:

$$\xi^2 \propto \xi_{AFM}^{6-d} \quad \text{for} \quad |q| \gg \xi_{AFM}^{-1}$$
$$\xi^2 \propto \xi_{AFM}^{d-2} \quad \text{for} \quad |q| \ll \xi_{AFM}^{-1}$$

Thus, for anisotropic three dimensional systems, it follows that a larger magnetic correlation length also gives rise to a larger nematic correlation length. Consequently, the established phase competition between superconductivity and magnetism will naturally shorten the magnetic correlation length, and given the above relation, also the nematic correlation length. This explains the sharp maximum of the nematic correlation length shown in Figure 4(e) and naturally resolves the apparent discrepancy between the phonon renormalization and the behavior of $C_{66}(0)$ in the superconducting state.

In summary, we have shown that TA phonons, which represent the soft mode of the nematic/structural phase transition in underdoped Ba(Fe$_{1-x}$Co$_x$)$_2$As$_2$ ($x < 0.06$), are renormalized by nematic fluctuations in optimally doped Ba(Fe$_{0.94}$Co$_{0.06}$)$_2$As$_2$. From the emergence of the non-linearity in the phonon dispersion, we were able to quantitatively determine the temperature dependence of the nematic correlation length $\xi$. $\xi$ peaks at the superconducting transition temperature, reaching the large value of 100 Å, i.e. about 40 times the Fe-Fe bond length. The temperature dependence of $\xi(T)$ is fully consistent with a nematic state competing with superconductivity. It is also consistent with the scenario in which nematic fluctuations arise from magnetic fluctuations, as the latter are known to compete with superconductivity. More generally, our work demonstrates the unique capabilities of INS, to probe finite-momentum nematic fluctuations, which are not accessible by the usual probes employed to study nematic fluctuations. A systematic study of how the nematic correlation length evolves with doping and chemical composition can potentially shed new light on the relevance of nematic fluctuations for superconductivity.

**Acknowledgments**
F.W. was supported by the Helmholtz Young Investigator Group under contract VH-NG-840. M.M. was supported by the Karlsruhe Nano-Micro Facility (KNMF). D.R. was supported by the DOE, Office of Basic Energy Sciences, Office of Science, under Contract No. DE-SC0006939. R.M.F. and M.S. are supported by the U.S. Department of Energy, Office of Basic Energy Sciences, under Award No. DE-SC0012336.

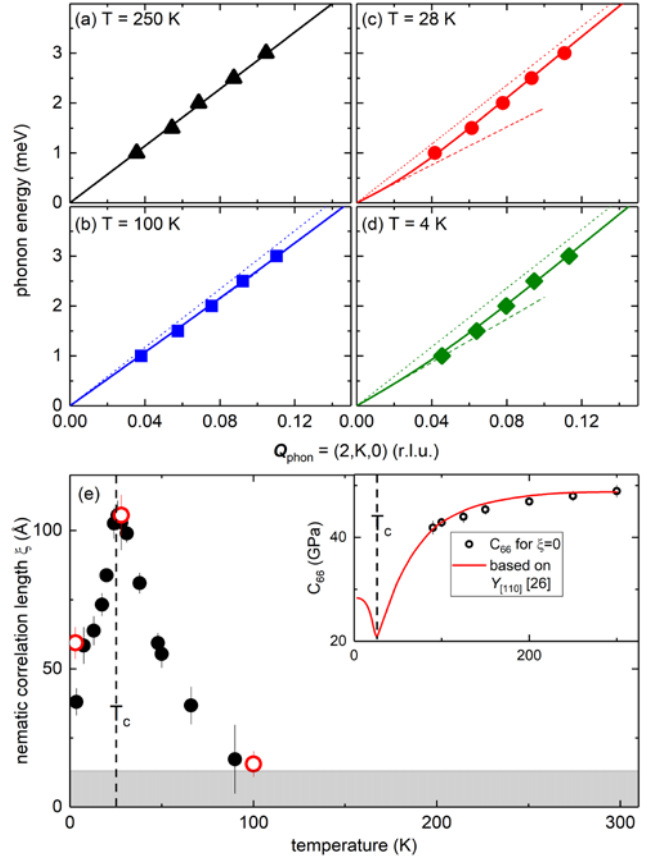

**FIG. 4.** (a)-(d) Analysis of the observed phonon dispersion (symbols) using Eq. (4) (solid lines) at four temperatures: two well above ($T = 100$ K, 90 K), one close to ($T = 28$ K), and one well below $T_c = 25$ K ($T = 4$ K). Dashed lines represent the linear component of the approximated dispersions at $q = 0$ ($E_{lin}(q \approx 0) = \sqrt{C_{66}(T) \cdot q^2/\rho}$, visible only for $T \leq 28$ K). Dotted lines represent the linear component of the estimated dispersion in the absence of nematic fluctuations ($E_{linear}(q \gg 0) = \sqrt{C_{66}^0(T) \cdot q^2/\rho}$). Error bars in $K$ of the experimental data are smaller than the symbol size. (e) Temperature dependence of the nematic correlation length $\xi$ deduced from the fits shown in (a)-(d) (red circles). The full dots represent results from an analysis with only three data points, i.e., taken at constant energies of E = 1.0 meV, 1.5 meV and 2.5 meV. The grey area indicates the limit below which we can't resolve $\xi$ anymore. The inset shows $C_{66}(T)$ (circles) deduced from linear fits ($\xi = 0$) to the phonon dispersion data for T $\geq$ 90 K. The red line was obtained by scaling the reported behavior of the Young's modulus $Y_{[110]}$ [26] to these measurements.

## Appendix

**Experimental setup**

Phonons were investigated using inelastic neutron scattering on the triple-axis spectrometers 4F2 and 1T located at the Laboratoire Léon Brillouin, CEA Sacly. We employed doubly focusing PG002 monochromators and analyers. All measurements were done in the [100]-[010] scattering plane. Measurements for energy transfers of 1.0 meV and 1.5 meV were performed at the 4F2 TAS with a fixed final energy of 8 meV. Measurements at higher energy transfers were performed on the 1T TAS with a fixed final energy of 14.7 meV.

The sample was one piece of single crystal of Ba(Fe$_{0.94}$Co$_{0.06}$)$_2$As$_2$ weighing 1 g and having a superconducting transition temperature of $T_c$ = 25 K. The doping value of the growth batch of $x = 0.06$ was determined by energy-dispersive x-ray spectroscopy and single-crystal x-ray diffraction at described in detail in Ref. [32].

The rocking scan of the (200) Bragg peak has a full width at half maximum (FWHM) of 0.8° (equivalent to 0.028 r.l.u.) and is within the experimental resolution. The FWHM did not change outside the experimental error between room temperature and T = 4 K, i.e., we did not observe any indication for a tetragonal-to-orthorhombic phase transition.

**Inelastic neutron scattering at E = 1.5 meV and 2.5 meV**

We show the full temperature dependences of phonon wavevectors $Q_{\text{phon}}$ observed at energy transfers of $E$ = 1.5 meV and 2.5 meV in Figure 5. Qualitatively, results for both energy transfers show the same behavior as those presented in Figure 3 for $E$ = 1.0 meV: Softening on cooling is followed by a flat temperature dependence in the temperature range $T_c \leq T \leq 70$ K. On cooling into the superconducting phase we again observe a softening.

**Details of the analysis of the phonon dispersion relation**

In our analysis of the phonon dispersion relation [Figs. 4(a)-(d)] based on Eq.(4)

$$E(q) = \sqrt{\frac{C_{66}^0}{\rho\left(1 + \frac{\lambda^2 \chi_{nem}(q=0)}{C_{66}^0(1+\xi^2 q^2)}\right)}} q^2$$

we make use of the reported data of the Young's modulus $Y_{[110]}(T)$ [26] which is closely related to the shear modulus $C_{66}(T)$ and hence to the slope of the transverse acoustic phonon dispersion. In particular, we take the temperature dependent values of $C_{66}(T)$ and $C_{66}^0(T)$ from these measurements which define the linear slope of the phonon dispersion at $q = 0$ and intermediate wave vectors, respectively. In fact, we can rewrite Eq.(4) only in term of $C_{66}(T)$ and $C_{66}^0(T)$ using the relation

$$\lambda^2 \chi_{nem}(q=0) = C_{66}^0 \left(\frac{C_{66}^0}{C_{66}} - 1\right)$$

resulting in a modified version of Eq. (4):

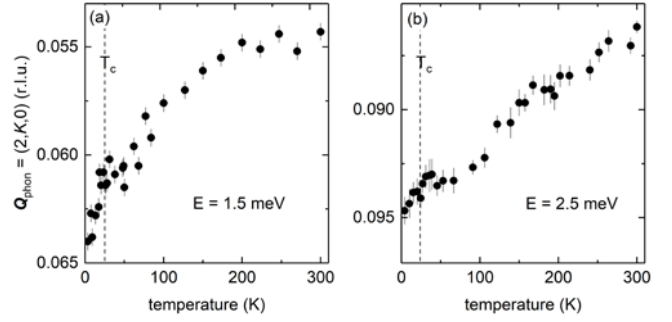

**FIG. 5.** Temperature dependence of phonon peak position at $Q_{phon} = (2, K, 0)$ (dots) observed at (a) $E$ = 1.5 meV and (b) 2.5 meV. For a more intuitive reading, we plot the $K$ scale upside-down because shift of the peak position to higher values of $K$ corresponds to a softening.

$$E(q) = \sqrt{\frac{C_{66}^0}{\rho\left(1 + \frac{C_{66}^0\left(\frac{C_{66}^0}{C_{66}}-1\right)}{C_{66}^0(1+\xi^2 q^2)}\right)}} q^2.$$

In the following, we want to explain how we deduced $C_{66}(T)$ and $C_{66}^0(T)$ from the reported measurements of the Young's modulus $Y_{[110]}(T)$ [26].

Since the Young's modulus $Y_{[110]}$ is nearly temperature-independent at 300K, we assume that nematic fluctuations are negligible at room temperature and, hence, $C_{66}(T = 300\text{ K}) = C_{66}^0(T = 300\text{ K})$. We can expect that the temperature dependence of $Y_{[110]}$ ($\propto C_{66}$) and that of the phonon energies to be the same as long as the phonons are in the linear dispersion regime with a slope corresponding to $C_{66}(T)$. This is the case for T ≥ 90 K where linear fits describe the phonon dispersions well, [see Figs. 4(a) and (b)]. Thus the slope of the dispersion obtained from linear fits to the data should match $Y_{[110]}$ ($\propto C_{66}$).

However, the phonon softening observed for T ≥ 90 K in inelastic neutron scattering is somewhat larger than the temperature dependence of $Y_{[110]}(T)/Y_{[110]}(300\text{K})$ suggests. This is clearly visible when we plot $Y_{[110]}/Y_{[110]}(T=300\text{K})$ only scaled to match the average high temperature value of 47.9 GPa [blue dashed line in Fig. 6(a)]. Here, $C_{66}$ softens to 63% of its room temperature value whereas scaling to the phonon softening observed in INS on cooling to $T$ = 90 K [circels in Fig. 6(a)] yields a bigger decrease down to 42% of the room temperature value. On the other hand, 3-point-bending experiments performed in order to determine $Y_{[110]}(T)$ put the sample under some non-negligible strain/pressure and, hence, the reported temperature dependence might be changed somewhat compared to the behavior in a non-strained sample. Physically, it makes no sense that the slope of the phonon dispersion at $q = 0$ results in larger phonon energies than we observe in neutron scattering at finite momenta.

Furthermore, it is instructive to discuss the relation between the Young's modulus $Y_{[110]}$ and $C_{66}$ in more detail. According to Ref. [38], Young's modulus along [110] can be written as

$$Y_{[110]} = 4\left(\frac{1}{C_{66}} + \frac{1}{\gamma}\right)^{-1} \quad \text{and} \quad \gamma = \frac{C_{11}}{2} + \frac{C_{12}}{2} - \frac{C_{13}^2}{C_{33}}.$$

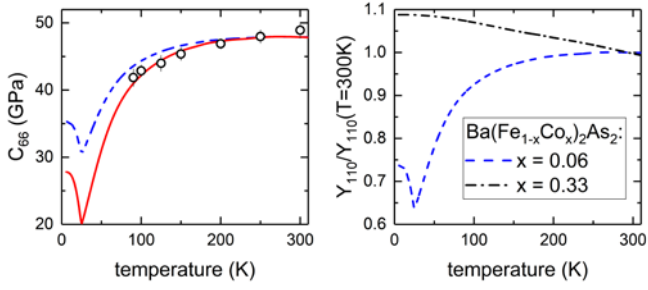

**FIG. 6.** (a) Temperature dependence of $C_{66}$ as deduced from linear fits to INS data for T ≥ 90 K [circles, same as in inset of Fig. 4(b)]. The dashed line represents the reported $T$ dependence of the Young's modulus $Y_{[110]}/Y_{[110]}(T=300K)$ [26] scaled to fit the high temperature value of 47.9 GPa observed in linear fits to our INS data for $T \geq 200$ K. The solid line [same as in inset of Fig. 4(e)] represents the $T$ dependence of $Y_{[110]}$ but scaled to fit the $T$ dependence of $C_{66}(T)$ deduced from linear fits, i.e., using $\xi = 0$, to the neutron data. (b) Young's modulus normalized to its value at T = 300K from three-point-bending experiments for Ba(Fe$_{1-x}$Co$_x$)$_2$As$_2$ with $x$ = 0.06 (blue dashed line) and 0.33 (balck dash-dotted line). Results are reproduced from Ref. [26].

Hence, $Y_{[110]}$ is dominated by $C_{66}$ as long as $C_{66}$ is smaller than the other $C_{ij}$. Even then, a close comparison of results from resonant ultrasound for $C_{66}$(T) [27] and three-point-bending measurements for $Y_{[110]}$(T) [26] reveals differences of more than 15% in similarly doped samples of Ba(Fe$_{1-x}$Co$_x$)$_2$As$_2$ [38]. Further and contrary to expectations, $Y_{[110]}$ does not reach zero at the structural phase transition in underdoped Ba(Fe$_{1-x}$Co$_x$)$_2$As$_2$, which may be related to effects of finite stress not present in neutron scattering measurements [26,38]. Hence, we believe that the differences in the curves shown in Figure 6(a) [circles and dashed line] are not intrinsic but rather caused by the experimental difficulties and approximations in the analysis of the published results.

Therefore, our analysis presented in Figure 4 is based on rescaled values of $C_{66}(T)$ shown as solid (red) line in Figure 6(a) [same as in the inset of Fig. 4(e)]. This approach is corroborated by the fact that our data for $T \geq 90$ K can be well approximated by straight lines. Regarding Eq. (4), this implies that the nematic correlation length $\xi(T)$ is so small at these temperatures that we do not see significant deviations from the linear behavior defined by the corresponding values of $C_{66}(T)$ in the investigated momentum range.

In order to have only one free parameter in our analysis, i.e., the nematic correlation length $\xi$, we estimate the slope of the expected linear dispersion in the absence of nematic fluctuation defined by $C_{66}^0(T)$ from the observed increase of the Young's modulus $Y_{[110]}$(T) on cooling in strongly overdoped Ba(Fe$_{0.67}$Co$_{0.33}$)$_2$As$_2$ similar to the procedure in Ref. [26]. In Ba(Fe$_{0.67}$Co$_{0.33}$)$_2$As$_2$ $Y_{[110]}$ increases by about 8% on cooling from room temperature to T = 5K [dash-dotted line in Fig. 6(b)]. This corresponds to a hardening of the respective phonon energies of about 4%. Such a hardening is typical for many solids and is directly related to the corresponding thermal expansion.

In our analysis the phonon dispersion is given by

$$E(q) = \sqrt{\frac{C_{66}^0}{\rho} q^2}$$

corresponds to the limiting case of Eq. (4) at large momenta. It is interesting to note that this limit is reached at smaller momenta for larger values of the nematic correlation length $\xi(T)$ [e.g. see Figs. 4(a)-(d)].